\documentclass[12pt]{article}

\usepackage[english]{babel}
\usepackage{float}
\usepackage{placeins}
\usepackage[letterpaper,top=2cm,bottom=2cm,left=3cm,right=3cm,marginparwidth=1.75cm]{geometry}
\usepackage{comment}
\usepackage{amsmath,amsfonts}
\usepackage{graphicx}
\usepackage{subcaption}
\usepackage[colorlinks=true, allcolors=blue]{hyperref}
\usepackage{authblk}

\def\frac#1#2{{\textstyle{#1\over#2}}}

\def\sign{{\rm sign}}

\def\endex{{\hfill{$\square$}\medskip}}

\newcommand{\D}[1]{\mathrm{d}{#1}}

\def\bi{\begin{itemize}}
\def\ei{\end{itemize}}
\def\bd{\begin{description}}
\def\ed{\end{description}}
\def\ben{\begin{enumerate}}
\def\een{\end{enumerate}}
\def\bv{\begin{verbatim}}
\def\ev\end{verbatim}

\def\s{{\sigma}}

\def\pr{{\rm Pr}}
\def\Pr{\pr}

\def\2to{{\ {\buildrel 2\over \longrightarrow}\ }}

\def\I1ton{{$I_1,\ldots,I_n$}}
\def\X1ton{{$X_1,\ldots,X_n$}}
\def\Y1ton{{$Y_1,\ldots,Y_n$}}
\def\Z1ton{{$Z_1,\ldots,Z_n$}}
\def\R1ton{{$R_1,\ldots,R_n$}}
\def\e1ton{{$e_1,\ldots,e_n$}}
\def\t1ton{{$t_1,\ldots,t_n$}}
\def\x1ton{{$x_1,\ldots,x_n$}}
\def\y1ton{{$y_1,\ldots,y_n$}}
\def\z1ton{{$z_1,\ldots,z_n$}}

\def\pvalue{{p-value}}
\def\pvalues{{p-values}}

\def\HBR{\textsc{hbr}}
\def\TCA{\textsc{tca}}

\title{Statistical Inference on the Miss Distance Compared to Collision Probability for Conjunction Analysis}

\author{Soumaya~Elkantassi\thanks{Postdoctoral Researcher, Department of Operations, University of Lausanne, Lausanne, Switzerland}, 
        Valérie~Chavez-Demoulin\thanks{Professor, Department of Operations, University of Lausanne, Lausanne, Switzerland}, 
        Anthony~C.~Davison\thanks{Professor, Institute of Mathematics, Ecole Polytechnique Fédérale de Lausanne, Lausanne, Switzerland}, 
        Matthew~D.~Hejduk\thanks{Senior Project Lead, Conjunction Assessment Program Office, The Aerospace Corporation, Chantilly, VA, USA},
        Hunter~A.~Morris\thanks{Member of the Technical Staff, Conjunction Assessment Program Office, The Aerospace Corporation, Chantilly, VA, USA}
        }

\date{}

\begin{document}
\maketitle

\begin{abstract}
Satellite conjunctions involving “near misses” of space objects are increasingly common, especially with the growth of satellite constellations and space debris. Accurate risk analysis for these events is essential to prevent collisions and manage space traffic. Traditional methods for assessing collision risk, such as calculating the so-called collision probability ($p_c$), are widely used but have limitations, including counterintuitive interpretations when uncertainty in the state vector is large. To address these limitations, we build on an alternative approach proposed by \cite{ElkantassiDavison:2022} that uses a statistical model allowing inference on the miss distance between two objects in the presence of nuisance parameters. This model provides significance probabilities for a null hypothesis that assumes a small miss distance and allows the construction of confidence intervals, leading to another interpretation of collision risk.

In this study, we compare this approach with the traditional use of $p_c$ across a large, NASA-provided dataset of real conjunctions, in order to evaluate its reliability and to refine the statistical framework to improve its suitability for operational decision-making.  We also discuss constraints that could limit the practical use of such alternative approaches.

\end{abstract}
\newpage
\section{Introduction}

The rapid expansion of satellite constellations and the persistent threat posed by orbital debris have greatly increased the frequency of close approaches in Earth orbit. To mitigate the risk of on-orbit collisions, satellite operators frequently monitor potentially hazardous conjunctions over time, refining the collision probability ($p_c$) as new tracking data become available.

Traditionally, $p_c$ is iteratively computed at each prediction update by integrating the density describing the relative position uncertainty over the disk defined by the sum of the colliding objects' radii. This iterative assessment is especially important because the positional uncertainties, and hence $p_c$, often evolve significantly between an initial detection of a close approach and the final decision point, typically just hours before the Time of Closest Approach (\TCA) \cite{Patera:2001,FosterEstes:1992,Chan:1997}. Meanwhile, operators can also track the evolution of the observed miss distance \cite{Alfano:1993}, an intuitive measure of how ``close" an encounter is likely to be. As each new piece of tracking data refines the orbital covariance, updated estimates of $p_c$ can either confirm a low-risk event or indicate that active avoidance maneuvers are warranted. However, relying solely on $p_c$ can lead to confusion when uncertainties are large. Under such circumstances, the so-called \emph{dilution phenomenon} manifests as artificially low values of $p_c$, creating a false sense of safety \cite{Balch:2019,Hejduk:2019}.

Recent work \cite{ElkantassiDavison:2022,Elkantasi:2023} proposes shifting the focus from $p_c$ to direct statistical inference on the miss distance at the \TCA. By framing conjunction assessment in terms of hypothesis testing, one obtains significance probabilities and confidence intervals that more transparently balance Type~I (false alarm) and Type~II (missed detection) errors. This perspective aims to offer decision-makers a clearer statement of the situation, even when orbital uncertainties are considerable. In this paper, we build on these ideas and compare the traditional approach of computing $p_c$ against the inference-based method for the miss distance using a large real-world dataset provided by NASA. We also show a theoretical relationship indicating that the conventional collision probability is systematically lower than the significance probability derived from hypothesis testing. 

The remainder of the paper is organized as follows. Section~\ref{sec:CA} reviews the conventional methodology for calculating collision probability and its known limitations. Section~\ref{sec:InferenceMissDistance} describes the inference framework built around the miss distance, including the relevant hypothesis tests and their interpretation. Section~\ref{sec:Results} presents the NASA conjunction dataset and the numerical comparison of both approaches. We conclude in Section~\ref{sec:conclusion} with a discussion of the operational implications of our findings and potential directions for future research.

\section{Collision Risk Assessment}\label{sec:CA}
\subsection{ Collision Probability $p_c$}
Accurate collision risk assessment is an essential aspect of space operations. A widely-used basis for risk assessment is the calculation of collision probability. This provides a quantitative estimate of the probability of two objects coming into contact during a close encounter and has become a standard basis for operational risk management.

The probability of collision, $p_c$, is derived within the following statistical framework. The observed state vector $y$, which represents a noisy estimate of the relative position $\mu$ and velocity $\nu$ of two objects, is assumed to follow a multivariate normal distribution with mean vector $\eta$ and known covariance matrix $\Omega$. The vector $y$ is the result of an orbit determination and prediction process, that is part of a broader space surveillance framework that relies on dynamic catalogs of space objects and comprehensive forecasting methods \cite{Chan:1997}. Like other approaches to conjunction analysis, our statistical formulation treats the state vector $y$ as a realization of a random variable drawn from the specified multivariate normal distribution.

In the case of short-term conjunctions, the following assumptions are taken to hold: (1) the encounter is brief, involving a single close approach at the Time of Closest Approach (\TCA); (2) the relative motion between the objects is approximately linear, with the relative velocity assumed constant; and (3) velocity uncertainty can be neglected, with all positional uncertainties projected onto the two-dimensional encounter plane normal to the relative velocity at the \TCA\ \cite{Hall:2021}.

\begin{figure}[h]
    \centering
    \includegraphics[width=0.6\linewidth]{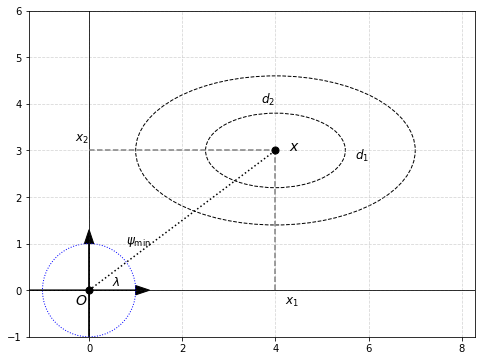}
   \caption{
        The geometry of an encounter in the plane normal to the relative velocity vector. 
        The observed position is $x = (4, 3)$, and its projections onto the two axes are shown by dashed black lines. The ellipses are centered at $x$, with standard deviations $d_1 = 1.5$ and $d_2 = 0.8$;  without loss of generality their semi-axes can be taken to be parallel to the coordinate axes. 
        The hard-body radius (\HBR), $\psi_{\text{min}} = 1$, shown as the blue dashed circle centered at the origin $O$, represents the collision volume $\mathcal{V}$. The angle $\lambda$ indicates the orientation of $x$ relative to the origin.
    } \label{fig:Figure1}
\end{figure}

These assumptions imply that the position uncertainties form a three-dimensional Gaussian error ellipsoid during the encounter, which yields a bivariate Gaussian distribution when projected onto the encounter plane. In this two-dimensional plane normal to $\nu$, the observed position $x$ is treated as a projection of $y$, modeled as a bivariate normal variable centered at the projected true position $\xi$, with covariance matrix $D = \text{diag}(d_1^2, d_2^2)$; see Figure~\ref{fig:Figure1}. The conjunction probability $p_c$ is then  computed as the integral of the joint normal probability density function $f(x; \xi)$ over the collision volume $\mathcal{V}$, defined as the disk of radius $\psi_{\text{min}}$ (the combined hard-body radius) centered at the origin, i.e., 
\begin{equation}
   p_c = \int_{\|\mathbf{x}'\| \leq \psi_{\text{min}}} f(x'; \xi) \, \D{d}x'. 
   \label{formula_Pc}
\end{equation}

Numerous methods to evaluate the integral in Equation~\eqref{formula_Pc} efficiently and accurately have been proposed. Early approaches relied on numerical quadrature techniques, while later methods leveraged analytical approximations and asymptotic expansions to reduce computational complexity. A widely used method \cite{FosterEstes:1992} employs a transformation to polar coordinates, simplifying the integral into a form suitable for numerical evaluation. Alternative approaches include the Rice integral method \cite{Patera:2001,Patera:2005,Patera:2003}, which relies on affine transformations and the ``line integral" method \cite{Alfano:2005}, which approximates the integral using contour integration. Analytical techniques such as Laplace transforms \cite{Serra_et_al:2016} and Hermite polynomial expansions \cite{Pelayo-Ayuso:2018} further refine probability estimates by leveraging closed-form approximations. A comparative analysis of these methods is available in \cite{Chan:2020}.

\subsection{Limitations of $p_c$}\label{Section:limitation of Pc}

\begin{figure}
    \centering
    \includegraphics[width=0.6\linewidth]{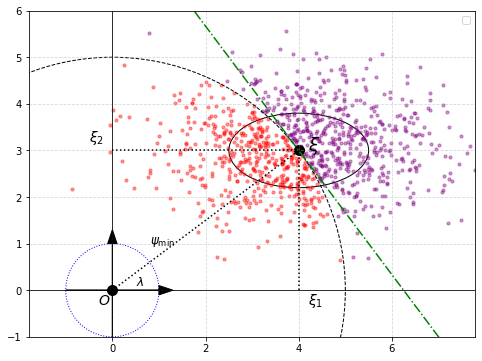}
 \caption{
        This Figure~corrects the geometry of Figure~\ref{fig:Figure1}. 
        The uncertainty is centered at the true relative position $\xi = (4, 3)$, with projections $\xi_1$ and $\xi_2$ onto the axes of the encounter plane. 
        The standard deviations are $d_1 = 1.5$ and $d_2 = 0.8$. 
        The hard-body radius $\psi_{\text{min}} = 1$, is illustrated as the blue dashed circle centered at the origin $O$, representing the collision volume. 
        1000 points were sampled from the bivariate normal distribution centered at $\xi$ with covariance $D = \text{diag}(d_1^2, d_2^2)$. 
        Points closer to $O$ than $\xi$ are shown in red, while points farther from $O$ are shown in purple. 
        The dashed black circle is the circle with radius $\psi$ passing through $\xi$, and the angle $\lambda$ indicates the orientation of $\xi$ relative to the origin.
    }
    \label{fig:Figure2}
\end{figure}

In the definition of the collision probability $p_c$ in equation~\eqref{formula_Pc}, the probability density function is centered at the true relative position $\xi$. However, since $\xi$ is unknown in practice, $p_c$ is estimated by centering the density at the observed position $x$, giving
\begin{equation*}
    \hat{p_c} = \int_{\|\mathbf{x}'\| \leq \psi_{\text{min}}} f(x'; x) \, \D{x'}.
\end{equation*}
One might think that computation of $\hat p_c$ is the basis for risk analysis, but unfortunately this is not the case, for two reasons.  The first is the underlying assumption that the conjunction geometry corresponds to Figure~\ref{fig:Figure1}, where the density is centered at the observed position $x$, whereas in reality the situation is better represented by Figure~\ref{fig:Figure2}, where the density is centered at the true relative position $\xi$. 

A bias arises because the observed position $x$, shown in Figure~\ref{fig:Figure2} as one of the sample points, incorporates measurement noise and is, on average, farther from the origin than the true position $\xi$. As a result, $\hat{p_c}$ tends to  underestimate the true collision probability $p_c$ by an amount that depends on the measurement uncertainty, increasing as the covariance matrix grows larger.

The distinction between $p_c$ and $\hat{p_c}$ is often overlooked. The plug-in estimate $\hat{p_c}$ is commonly referred to as $p_c$ without acknowledging the systematic bias introduced by centering the density at $x$ instead of $\xi$~\cite{ElkantassiDavison:2022, Elkantasi:2023}. This lack of clarity can lead to misinterpretation of the estimated collision probability, as $\hat{p_c}$ does not account for the true conjunction geometry but rather approximates the probability based on a single noisy observation. Clarifying this is essential for accurate and reliable risk assessments in operational contexts~\cite{Hall:2025}.

A more serious difficulty is that $p_c$ is the probability that the observation vector $x$, and not the true position $\xi$ of the satellite,  lies inside the hard-body radius.  In classical statistics $\xi$ is regarded as an unknown but fixed constant, so the latter probability is either 0 or 1.  Thus a different statistical approach is needed in which $\xi$ is treated as a random variable.  

A limitation of $\hat{p_c}$ is its sensitivity to uncertainties in the state vector, a phenomenon often called the dilution effect~\cite{Balch:2019,Hejduk:2019}. When the covariance matrix is small (i.e., uncertainty is low), the probability density function (PDF) becomes tightly concentrated around its mean, reducing the probability inside the collision volume and leading to a lower $\hat{p_c}$, whereas when the covariance matrix is large (i.e., uncertainty is high), the PDF spreads over a much broader area, again decreasing $\hat{p_c}$. Between these extremes, there is an intermediate level of uncertainty where the overlap between the PDF and the collision volume is maximized~\cite{Alfano:2005aa}.

This behavior highlights the sensitivity of $\hat{p_c}$ to the level of uncertainty and complicates its interpretation. In practice, some operational teams compute $\hat{p_c}_{\text{max}}$, the highest value of $\hat{p_c}$ across possible uncertainties, comparing it to fixed thresholds such as $10^{-3}$, $10^{-4}$, or $10^{-7}$ to guide mitigation decisions. The selection of these thresholds has evolved with little formal consideration of how they should be set.

\subsection{Alternative methods}

Methods developed to address collision risk assessment challenges include belief and plausibility-based approaches, such as decision support systems combining Dempster--Shafer theory with machine learning~\cite{Mellado_etal:2023}, which provide nuanced classifications based on collision probability, uncertainty and mitigation costs. Simulation studies suggest that this approach provides detailed risk assessments, although its routine operational implementation remains under evaluation.

Fully numerical methods, such as Monte Carlo simulations, and machine learning-based approaches~\cite{Tulczyjew:2021, Catulo_et:2023}, are also widely applied. While effective in many cases, these methods can be computationally burdensome.

 One approach focuses on the miss distance, either deriving its distribution \cite{Chan:2011,Carpenter:2019}, or inverting its moment-generating or characteristic functions in order to estimate collision probabilities \cite{Bernstein.etal:2021}. These approaches share similarities with the framework proposed in \cite{ElkantassiDavison:2022}, but the latter move beyond probability calculations to inference about the miss distance. The details of this approach are presented in the next section.

\section{Inference on the miss distance}
\label{sec:InferenceMissDistance}

The framework proposed in \cite{ElkantassiDavison:2022}, building upon earlier ideas in \cite{FosterEstes:1992,Chan:1997}, shifts the focus from estimating the collision probability to inference about the distance between the objects at the time of closest approach. Rather than using collision probability as the sole metric, this approach treats the true miss distance $ \psi $ as the key quantity of interest, enabling the construction of confidence intervals and significance probabilities that can yield a more informative assessment of conjunction risk.

\subsection{Statistical model}
\label{subsec:StatModel}

The underlying statistical model is that used above, i.e., the observed relative position $x$ is treated as a bivariate normal variable with mean $\xi$ and  covariance matrix $D = \mathrm{diag}(d_1^2, d_2^2)$ obtained by projecting the three-dimensional position covariance onto the encounter plane and rotating it as described in \cite{ElkantassiDavison:2022}. Our goal is to make inference about the length of $\xi$, which is expressed in polar coordinates as
\begin{equation}
  \xi
    =
  (\psi \cos \lambda,\;\psi \sin \lambda),
  \label{state_2D}
\end{equation}
where $\psi > 0$ is the Euclidean norm of $\xi$ (the miss distance) and $ \lambda \in [0,2\pi)$ is the orientation angle in the encounter plane. The probability density function of $x$ is thus
\begin{equation}
  f(x;\,\xi)
    =
  \dfrac{1}{2\pi\,d_1\,d_2}
  \,\exp\!\left[
    -\tfrac{1}{2}\left\{
       \dfrac{(x_1 - \psi \cos \lambda)^2}{d_1^2}
      +\dfrac{(x_2 - \psi \sin \lambda)^2}{d_2^2}
    \right\}
  \right].
  \label{density_2D}
\end{equation}

When the relative motion is linear but velocity uncertainty must be incorporated, the observed state becomes a six-dimensional vector $y$, following
$$
  y \sim \mathcal{N}_6(\eta, \Omega),
$$
with $\eta$ and $\Omega$ respectively denoting the mean and covariance of the full state. The density is then
$$
  f(y;\,\eta)
    =
  \dfrac{1}{(2\pi)^3 \,|\Omega|^{1/2}}
  \,\exp\!\left\{
    -\dfrac12\,(y - \eta)^T\,\Omega^{-1}\,(y - \eta)
  \right\}.
$$
In this higher-dimensional setting, the miss distance $\psi$ depends on both the mean relative position $ \mu $ and the velocity vector $\nu$. Rather than using $\|\mu\|$ directly, we write $
  \psi = \|\mu\|\;\bigl|\sin \beta\bigr|$, 
where $\beta$ is the angle between $\mu$ and $\nu$. Accounting for both position and velocity uncertainties raises the dimension of the parameter space, yet $\psi$ remains the parameter of primary interest, with additional angles and orientation parameters treated as nuisance components. In the two-dimensional case, $\lambda$ describes the inclination of $\xi$ in the plane, while in the six-dimensional scenario, $\lambda$ generalizes to include the unknown orientation of the encounter plane and alignment of  $\mu$ and $\nu$.

\subsection{Hypothesis formulation}
\label{subsec:HypothesisFormulation}

Statistical inference aims to draw conclusions about unknown parameters based on the observed data. In parametric settings such as those above a central quantity is the likelihood function, defined as the probability density function of the observed data, regarded as a function of the parameters.  The likelihood, which is a central focus of interest in both frequentist and Bayesian approaches to inference, typically depends on both a parameter of interest $\psi$ and nuisance parameters $\lambda$. A common strategy to isolate $\psi$ uses for example the profile log-likelihood function \cite{Davison:2003}
$$
  \ell_p(\psi)
    =
  \ell\bigl\{\psi,\;\hat{\lambda}(\psi)\bigr\},
  \quad
  \hat{\lambda}(\psi)
    =
  \arg\max_\lambda
  \,\ell(\psi,\lambda)
$$
as the basis for pivotal statistics such as the \emph{likelihood root}\ and \emph{Wald}\ statistics
\begin{flalign}
  r(\psi_0)
    &= \mathrm{sign}\bigl(\hat{\psi}-\psi_0\bigr)\;\sqrt{
         2\left\{\ell_p(\hat{\psi})-\ell_p(\psi_0)\right\}
       },
    \label{eq:likelihood_root}
  \\
  w(\psi_0)
    &= \dfrac{\bigl(\hat{\psi}-\psi_0\bigr)^2}%
            {\widehat{\mathrm{var}}(\hat{\psi})},
    \label{eq:wald_statistic}
\end{flalign}
where $\widehat{\mathrm{var}}(\hat{\psi})$ is typically derived from the observed information matrix. Under regularity conditions and in sufficiently large samples, and if $\psi_0$ is the true value of $\psi$, both $r(\psi_0)$ and $w(\psi_0)$ have approximate standard normal distributions, making them bases for hypothesis tests and confidence intervals despite the presence of the nuisance parameters.
The resulting inferences are said to be well-calibrated if the standard normal distributional approximations for $r(\psi_0)$ and $w(\psi_0)$ are accurate, in which case tests and hypothesis tests based on them will have good frequentist properties.  In the present setting there is a single observation, and the `large sample size' corresponds to small variance, i.e., small $d_1$ and $d_2$ in the two-dimensional setting, or small $\Omega$ in a suitable matrix sense in six dimensions.  

The likelihood framework sketched above is a well-established basis for statistical inference that is used in a vast range of applications. In the context of conjunction assessment, these principles are applied to ensure robust risk evaluation regarding close approaches. One key challenge is to decide whether such an approach requires mitigation, and this can be addressed via a hypothesis test of whether the miss distance $ \psi $ exceeds a predefined safety threshold $\psi_0$. We might test 
\begin{equation}
H_0: \psi \geq \psi_0 
\quad \text{vs.} \quad 
H_A: \psi < \psi_0, 
\label{test_setup1}
\end{equation}
using the smallest significance probability under $H_0$, i.e., 
\begin{equation}
    p_{\rm obs} = \Pr\bigl\{r(\psi_0) \geq r_{\rm obs}\bigr\} = \Phi\bigl(-r_{\rm obs}\bigr), \label{p-obs}
\end{equation}
where $\Phi$ denotes the standard normal cumulative distribution function, and reject the null hypothesis $H_0$ if $p_{\rm obs}$ falls below a chosen significance level $\alpha$.  

The setup in~\eqref{test_setup1} allows for well-calibrated inference and ensures that both over- and under-estimation risks are accounted for. Even with large uncertainty, framing the test in this manner can yield reliable inferences by balancing the probabilities of Type~I and Type~II errors.  In the present context a Type~I error corresponds to an unnecessary mitigation maneuver, a \emph{false alarm},\ whereas a Type~II error corresponds to not making such a maneuver when one is necessary to avoid a collision, a \emph{missed detection}.\  Rejecting $H_0$ corresponds to identifying a potentially critical conjunction. This helps avoid frequent false alarms but can understate the risk when uncertainty is large, thus increasing the chance of missed detections \cite{Hejduk:2019}.


The choice of significance level $\alpha$ therefore involves a trade-off between Type~I and Type~II errors. Smaller $\alpha$ decreases the rate of false alarms but raises the rate of missed detections, and this trade-off grows more severe with high uncertainty. A complementary approach is to use confidence intervals, which provide a range of plausible values for 
$\psi$ rather than a binary accept/reject decision. Confidence intervals, particularly those derived from profile-likelihood methods, can offer a more flexible and informative approach to conjunction assessment by directly incorporating uncertainty into  inferences \cite{Reid:1999,Greenland:2016aa}.

In the next section we illustrate this tradeoff using real conjunction data. In forming this tradeoff, it is essential to consider the underlying theoretical principles that separate the metrics of $p_c$ and $p_{\text{obs}}$. A detailed derivation of the relationship between these two metrics is presented in the Appendix~\ref{appendix:pc_vs_pvalues}, which shows, $p_c$ in Eq~\eqref{formula_Pc} is always lower than the significance probability in Eq~\eqref{p-obs}. This formal comparison underscores the conceptual differences in directly comparing their values.

\section{Analysis and Results}\label{sec:Results}
\subsection{Data description}\label{Data_description}
The dataset analyzed in this study is provided by NASA and consists of detailed records of close encounters between space objects. Each record provides the positional and velocity vectors of the primary and secondary objects, expressed in both Earth-Centered Inertial (ECI) and UVW coordinate systems. The dataset also includes covariance matrices that quantify the uncertainties associated with these state vectors, incorporating both three-dimensional positional and velocity covariances, along with their projections onto the encounter plane.

The dataset includes temporal information such as Conjunction Data Message creation time and Time to Closest Approach \TCA. Events are identified by grouping entries with the same primary and secondary object IDs within a $\pm 15$-minute window around the \TCA. To ensure sufficient time for collision avoidance, only the entries closest to 12 hours before \TCA\ are retained. After filtering, 76,225 unique OCMs remain, representing 9.38\% of the original 812,632 entries, forming the basis for the comparative analysis.

\begin{figure}[h]
    \centering
    \begin{minipage}{0.49\textwidth}
        \centering
        \includegraphics[width=\linewidth]{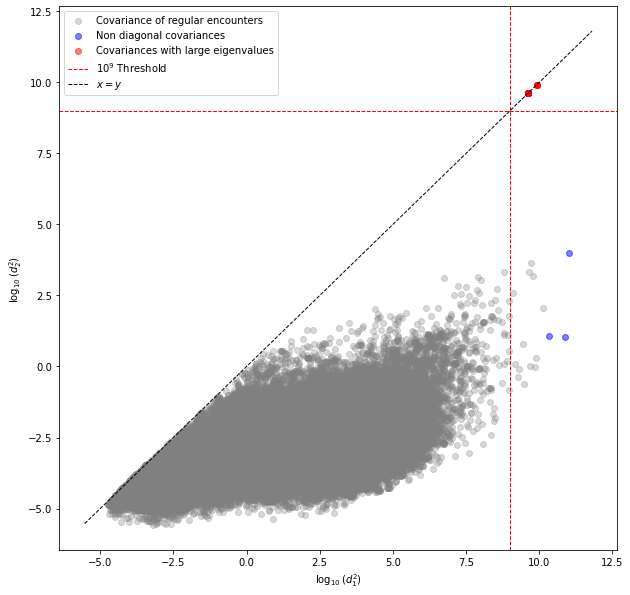}
    \end{minipage}
    \hfill
    \begin{minipage}{0.49\textwidth}
        \centering
        \includegraphics[width=\linewidth]{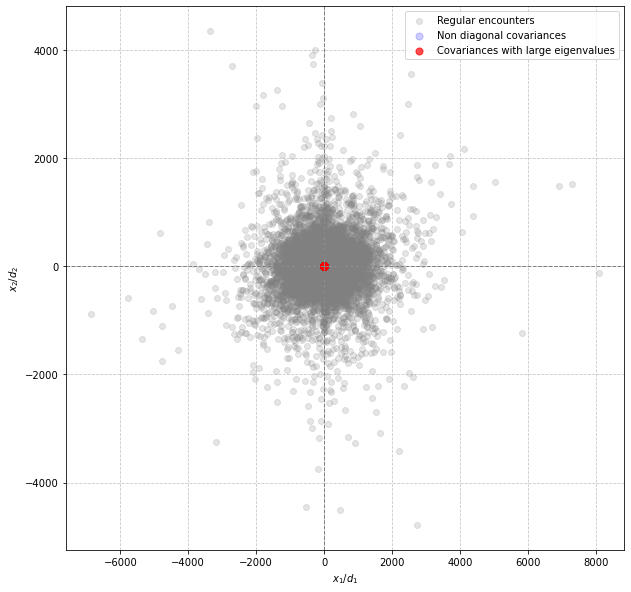}
    \end{minipage}
    \caption{Left: Scatter plot of log-scaled squared eigenvalues $ d_1^2 $ vs. $ d_2^2 $. Right: Scatter plot of standardized position variables $ x_1/d_1 $ vs. $ x_2/d_2 $.}\label{fig:scatter_d1d2&standardized_x1x2}
\end{figure}

Figure~\ref{fig:scatter_d1d2&standardized_x1x2} illustrates key spatial and uncertainty properties of the dataset. The left-hand panel shows a log-scaled scatter plot of the squared eigenvalues $d_1^2$ versus $d_2^2$, highlighting strong anisotropy in uncertainty. The eigenvalues often differ by several orders of magnitude, leading to highly elongated error ellipses, and, in extreme cases, nearly degenerate shapes resembling lines. These anisotropies significantly impact collision probability estimates and the interpretation of encounter dynamics. The right-hand panel of Figure~\ref{fig:scatter_d1d2&standardized_x1x2} shows standardized position variables $x_1/d_1$ versus $x_2/d_2$, with most encounters clustering near the origin. This occurs either because uncertainties are well-contained relative to position or because they are disproportionately large. A few outliers with extreme standardized values correspond to cases where uncertainties are small relative to positional coordinates in one or both directions.

\begin{figure}[h]
    \centering
    \includegraphics[width=\linewidth]{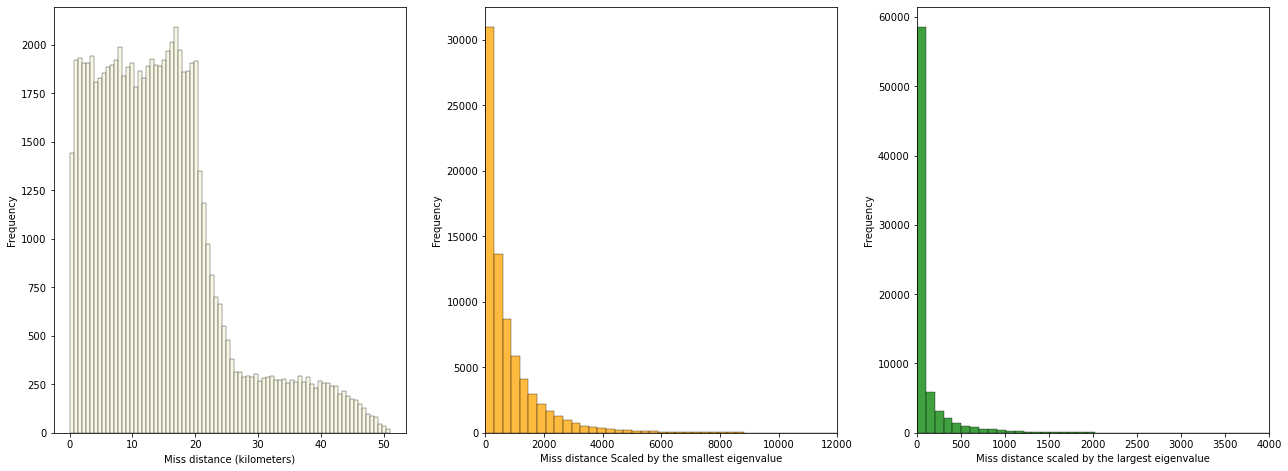}
    \caption{Histograms of estimated miss distances. 
Left: Distribution of absolute estimated miss distances, showing a concentration of encounters within 20 km. 
Middle: Estimated miss distances normalized by the smallest eigenvalue, exhibiting a heavy-tailed distribution.
Right: Estimated miss distances normalized by the largest eigenvalue, showing a more compressed distribution.}
\label{fig:histograms_miss_distance}
\end{figure}

A complementary view of encounter geometry and uncertainty is given in Figure~\ref{fig:histograms_miss_distance}, which shows the distribution of miss distances in absolute and normalized terms. The left-hand panel displays absolute miss distances, with most encounters occurring within 20 km. The middle and right-hand panels scale miss distances by the smallest and largest eigenvalues, respectively, and show a sharper peak near zero. This suggests that many encounters have small miss distances relative to uncertainty in at least one principal direction, an effect amplified when scaling by the largest eigenvalue, reinforcing the strong anisotropy in conjunction uncertainties.

The dataset includes estimates of $p_c$ using the method of \cite{FosterEstes:1992}, evaluated with both a standard hard-body radius (\HBR) of 20 meters and a mission-specific \HBR\ (Foster 1 and Foster 2), and the method of \cite{Chan:1997}, evaluated for the same two values of \HBR\ (Chan 1 and Chan 2). Table~\ref{tab:p_c} summarizes the resulting values of $\hat p_c$, which cluster near zero, as indicated by a $75\%$ quantile of essentially zero. However, higher probabilities are observed for methods using \HBR$ = 20$m, leading to more encounters exceeding the $(10^{-7}, 10^{-4})$ thresholds. A small fraction of the catalogued encounters, around $0.7\%,$ of the events reach values that could warrant maneuver planning.

\begin{table}[h]
\centering
\small 
\begin{tabular}{lcccc}
\hline
\textbf{Statistic} & $\hat{p}_c$ Foster 1 & $\hat{p}_c$ Foster 2 & $\hat{p}_c$ Chan 1 & $\hat{p}_c$ Chan 2 \\
\hline
Mean               & 3.1359e-05  & 1.7762e-05  & 3.1885e-05  & 1.8220e-05  \\
Median            & 7.4835e-188 & 3.7307e-193 & 8.8650e-199 & 1.7008e-199 \\
Min               & 0.0000      & 0.0000      & 0.0000      & 0.0000      \\
Max               & 3.4777e-01  & 3.4777e-01  & 3.4299e-01  & 3.4137e-01  \\
25\% Quantile      & 0.0000      & 0.0000      & 0.0000      & 0.0000      \\
50\% Quantile     & 7.4835e-188 & 3.7307e-193 & 8.8650e-199 & 1.7008e-199 \\
75\% Quantile      & 2.4259e-25  & 7.5545e-27  & 6.3523e-27  & 1.0519e-27  \\
90\% Quantile     & 1.7758e-07  & 1.6612e-08  & 1.1622e-07  & 1.3268e-08  \\
95\% Quantile     & 1.3546e-05  & 1.6050e-06  & 1.1919e-05  & 1.4312e-06  \\
99\% Quantile     & 2.6485e-04  & 8.4446e-05  & 2.4651e-04  & 7.4343e-05  \\
\hline
Encounters with $\hat{p}_c > 10^{-7}$ & 10568 (13.86\%) & 7045 (9.24\%) & 10419 (13.67\%) & 6949 (9.12\%) \\
Encounters with $\hat{p}_c > 10^{-4}$ & 503 (0.66\%) & 318 (0.42\%) & 493 (0.65\%) & 268 (0.35\%) \\
\hline
\end{tabular}
\caption{Summary statistics for the estimated collision probabilities $\hat p_c$, including the numbers exceeding critical probability thresholds.} 
\label{tab:p_c}
\end{table}

\subsection{Results and Discussion}

Although both $\hat{p}_c$ and p-values provide insights into conjunction risk, they assess it in fundamentally different ways. $\hat{p}_c$ estimates the probability of collision by integrating the joint density over the collision region, while $p_{\rm obs}$ measures the consistency of the observed data with a default assumption that a collision will occur, corresponding to the null hypothesis $H_0: \psi = \psi_0$, where $\psi_0$ is typically set to a safety threshold. Given these differences, their numerical values are not comparable. Moreover, as shown in the appendix, $\hat p_c \leq p_{\rm obs}$. This further emphasizes the  distinction between them and the resulting challenges of direct numerical comparison, and implies that $\hat p_c$ is not calibrated.

Beyond these theoretical differences, $p_c$ also has well-documented limitations as a measure of collision risk, discussed in Section~\ref{Section:limitation of Pc}. These limitations suggest that it should not be treated as an absolute benchmark for assessing conjunction risk. However, given its widespread use in operational conjunction assessments, its familiarity among practitioners makes it a practical reference when introducing alternatives. 

\begin{figure*}[ht]
    \centering
    \begin{subfigure}[b]{0.32\textwidth}
        \centering
        \includegraphics[width=\linewidth]{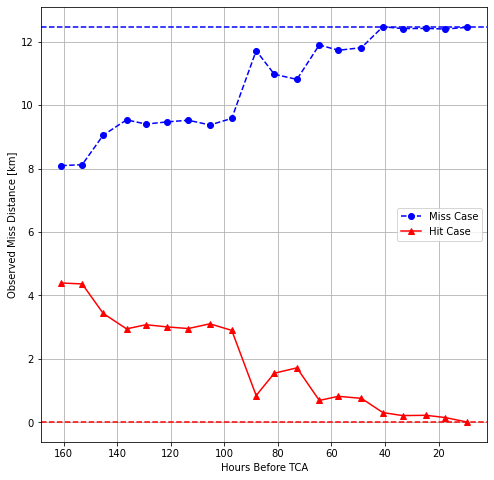}
        \label{fig:miss_distance_hit_miss}
    \end{subfigure}
    \hfill
    \begin{subfigure}[b]{0.32\textwidth}
        \centering
        \includegraphics[width=\linewidth]{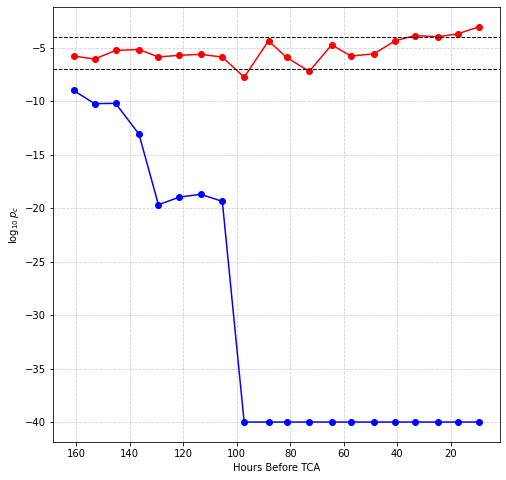}
        \label{fig:pc_hit_miss}
    \end{subfigure}
    \hfill
    \begin{subfigure}[b]{0.32\textwidth}
        \centering
        \includegraphics[width=\linewidth]{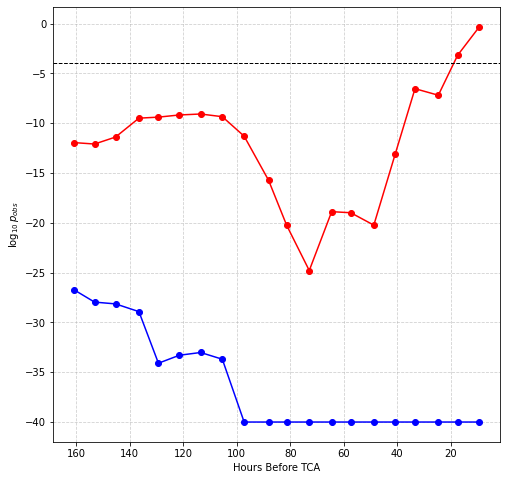}
        \label{fig:pobs_hit_miss}
    \end{subfigure}
    \caption{Left: observed miss distance for a particular conjunction in hit (red) and miss (blue) cases, where the hit trajectory leads to a collision while the miss trajectory does not. Middle and right: corresponding evolutions of $\log_{10} \hat{p}_c$ and $\log_{10} p_{\rm obs}$.}
    \label{fig:comparison_hit_miss}
\end{figure*}

In this section, we take as a reference $\hat p_c$ computed using the method in \cite{FosterEstes:1992} with a mission-specific \HBR, ensuring consistency with the \pvalues, which are also evaluated at $\psi_0 = \text{\HBR}$. To illustrate the behavior of $\hat p_c$ and \pvalue~over time, we construct a hypothetical scenario based on a single conjunction with multiple CDMs chosen from the original dataset. Initially, a nominal trajectory is considered, corresponding to the \textit{miss case}, where the object follows its original path. A second trajectory, the \textit{hit case}, is generated by biasing the state vector by $-\xi(\TCA)$ and thus shifting the trajectory toward a collision at the final recorded entry. Both trajectories are further perturbed by adding noise proportional to the final covariance, ensuring that the perturbation remains small. 
Figure~\ref{fig:comparison_hit_miss} presents the time evolution of the observed miss distance (left), $\hat p_c$ (middle), and $p_{\rm obs}$~(right) for this conjunction. As expected, the hit case (red) drifts toward zero miss distance, while the miss case (blue) remains at a larger separation. The values of $\hat p_c$ and $p_{\rm obs}$ are initially close for both cases but diverge significantly in the final 48 hours, with the hit case exceeding the $10^{-4}$ threshold. Figure~\ref{fig:comparison_hit_miss} demonstrates that as the uncertainties decrease sufficiently to yield a clear decision, both the \pvalue and $\hat{p}_c$ converge to either very small or very large values. However, during this process, the two metrics differ in absolute magnitude; as indicated by the differing scales on the y-axis. A more detailed Receiver Operating Characteristic (ROC) study, which examines the performance of $p_{\rm obs}$ and $\hat{p}_c$ over a range of hit–miss separations using similar data, is presented in \cite{Carpenter:2025aa}.

\begin{figure}[h]
    \centering
    \begin{minipage}{0.49\textwidth}
        \centering
        \includegraphics[width=\linewidth]{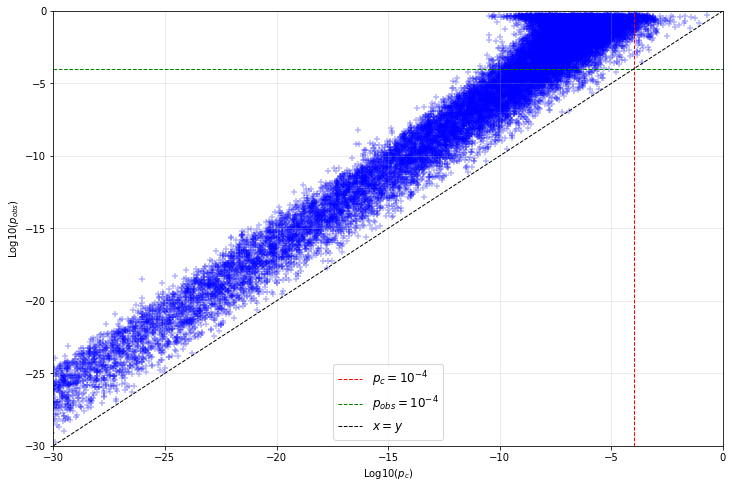}
    \end{minipage}
    \hfill
    \begin{minipage}{0.49\textwidth}
        \centering
        \includegraphics[width=\linewidth]{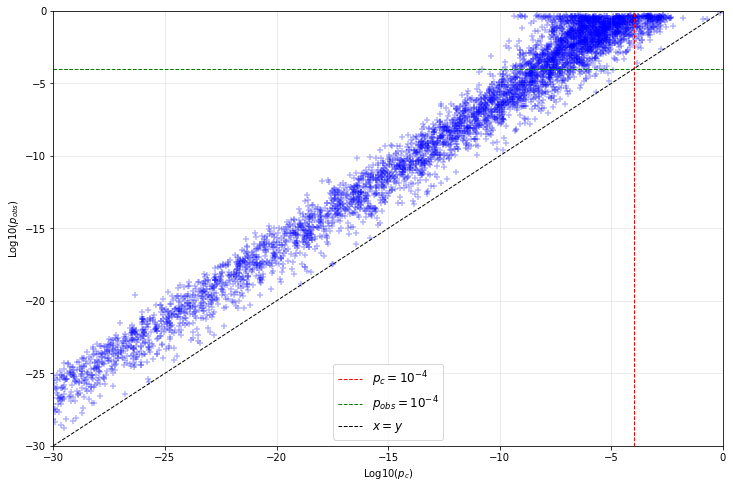}
    \end{minipage}
    \caption{Comparison of $\log_{10} \hat p_c$ and $\log_{10} p_{\rm obs}$ for different covariance scaling factors. The left panel corresponds to $ c = 1 $, representing results obtained using the original dataset, while the right panel shows results when all covariance matrices are scaled down by a factor of 100, i.e., $ c = 0.01 $. The dashed black line represents $ x = y $. The red dashed lines mark the threshold $ 10^{-4} $ for both metrics.}
\label{fig:profiling}
\end{figure}

The alignment observed near the \TCA~in the preceding example is not guaranteed, as we see from the right-hand panel of Figure~\ref{fig:profiling}, which shows values of $\hat{p}_c$ and $p_{\rm obs}$ for the dataset of conjunctions described in Section~\ref{Data_description}.  The values of $p_{\rm obs}$ are all larger than the corresponding values of $\hat p_c$, but while both metrics lead to the same classification outcome for 83\% of encounters, the discrepancies become more pronounced at higher values of $p_{\rm obs}$. 

This is also seen in the confusion matrix, which classifies encounters based on whether $p_{\rm obs}$ exceeds a threshold $\alpha$, when compared to the $10^{-4}$ threshold for $p_c$. The top left of Table~\ref{tab:confusion_matrices} confirms that $p_{\rm obs}$ gives a higher false alarm rate than $\hat p_c$, with significantly more encounters surpassing the $10^{-4}$ threshold but no missed detections. When the threshold for $p_{\rm obs}$ is increased to $\varepsilon=10^{-1}$ (top right), the false alarm rate decreases, but 79 encounters are then classified as missed detections. This illustrates how $p_{\text{obs}}$ naturally flags more encounters as high risk than $\hat p_c$.  Adjusting the threshold offers a controlled way to balance false alarms against missed detections.

A key factor influencing these results is the level of uncertainty. Shrinking the covariance by a factor $c = 0.01$ reduces the values of $\hat p_c$ and shifts many encounters below the $10^{-4}$ threshold, as shown in the right-hand panel of Figure~\ref{fig:profiling}; it lowers the false alarm rate, as reflected in the bottom matrices of Table~\ref{tab:confusion_matrices}. This underscores the strong dependence of both $\hat p_c$ and $p_{\rm obs}$ on uncertainty. However, interpreting false alarm and missed detection rates requires caution. Labeling a discrepancy as a ``missed detection'' presupposes that $\hat p_c$ provides the correct risk assessment, despite its known limitations. Since no actual collisions are recorded in the dataset, a missed detection does not imply a failure to predict a real collision but rather a classification divergence between $\hat p_c$ and p-values. If a real collision had occurred, failing to flag that event as high risk would be an obvious mistake. However, in the absence of recorded collisions, what we call a missed detection is just a case where $p_{\rm obs}$ and $\hat p_c$ disagree.
The work in \cite{Carpenter:2025aa} shows that $p_{\rm obs}$, computed based on the likelihood root, consistently provides superior discrimination between collision (hit) and non-collision (miss) cases when the data is constructed with a balanced $50\%$ of hits and $50\%$ of misses.
False alarms are more nuanced, occurring when $p_{\rm obs}$ flags an event for closer scrutiny even though $\hat p_c$ does not classify it as high risk. Unlike standard classification problems where false positives can be validated against real outcomes, here there is no definitive ground truth for collision probability. Since both metrics measure risk differently, their discrepancies do not necessarily indicate errors but rather variations in how uncertainty affects the outcomes. The work in \cite{Carpenter:2025aa} shows that $p_{\rm obs}$, computed based on the likelihood root, consistently provides superior discrimination between collision (hit) and non-collision (miss) cases when the data is constructed with a balanced $50\%$ of hits and $50\%$ of misses.

\begin{table*}[ht]
    \centering
    \begin{tabular}{lcc|cc}
        \hline
        \multicolumn{5}{c}{Original data} \\
        \hline
        & \multicolumn{2}{c|}{$\alpha = 10^{-4}$} & \multicolumn{2}{c}{$\alpha = 10^{-1}$} \\
        & $ \hat{p}_c < 10^{-4} $ & $ \hat{p}_c \geq 10^{-4} $ & $ \hat{p}_c < 10^{-4} $ & $ \hat{p}_c \geq 10^{-4} $ \\
        \hline
        $ p_{\rm obs} \geq \alpha $ & 11261 & 315 & 5010 & 236 \\
        $ p_{\rm obs} < \alpha $ & 63568 & 0 & 69819 & 79 \\
        \hline
    \end{tabular}

    \vspace{0.1cm} 

    \begin{tabular}{lcc|cc}
        \hline
        \multicolumn{5}{c}{Data with covariance matrices shrunk by $c=0.01$.} \\
        \hline
        & \multicolumn{2}{c|}{$\alpha = 10^{-4}$} & \multicolumn{2}{c}{$\alpha = 10^{-1}$} \\
        & $ \hat{p}_c < 10^{-4} $ & $ \hat{p}_c \geq 10^{-4} $ & $ \hat{p}_c < 10^{-4} $ & $ \hat{p}_c \geq 10^{-4} $ \\
        \hline
        $ p_{\rm obs} \geq \alpha $ & 1594 & 197 & 443 & 141 \\
        $ p_{\rm obs} < \alpha $ & 73410 & 0 & 74561 & 56 \\
        \hline
    \end{tabular}
    
    \caption{Confusion matrices comparing $\hat{p}_c $ and $ p_{\rm obs} $ for the original dataset (top) and the dataset with shrunk covariances (bottom) for significance levels $\alpha=10^{-4}, 10^{-1}$.}
    \label{tab:confusion_matrices}
\end{table*}
\section{Conclusion}\label{sec:conclusion}
In this paper, we formulated satellite conjunction assessment in statistical terms and introduced an inference framework centered on the miss distance. Our previous analyses \cite{ElkantassiDavison:2022,Elkantasi:2023} demonstrated that the conventional estimator, $\hat{p}_c$, is downward biased because it relies on plug-in estimates and fails to capture the true conjunction geometry. Here, we further showed that the collision probability is always lower than the significance probability $p_{\rm obs}$ derived from the signed likelihood root.

Using a large, real conjunction dataset provided by NASA, we compared the collision probability estimates with the significance probabilities for conjunction assessment. Our numerical results indicate that $p_{\rm obs}$ produces a higher false alarm rate than $\hat{p}_c$; moreover, this false alarm rate decreases as the significance level is lowered and uncertainty is reduced. This is primarily attributable to the construction of the null hypothesis; when the observed data for a particular conjunction lack sufficient precision, as is often the case many hours before \TCA, there is stronger evidence in favor of the null.

The comparison between these two metrics naturally encompasses the discussion of false alarms and missed detections. Since $\hat{p}_c$ is computed as an integral estimate and does not inherently control error rates, thresholds based on it do not provide the calibrated trade-off between Type I and Type II errors that the hypothesis testing framework underlying $p_{\rm obs}$ offers. Consequently, while thresholds applied to $\hat{p}_c$ can guide decision-making, they do not offer the same level of error rate control as a formal hypothesis test.

Future work should explore further refinements to improve the operational suitability of our statistical approach, particularly in managing false alarms, selecting appropriate significance thresholds, and incorporating time series information from repeated observations of the same encounter.
\section*{Acknowledgements}

 This material is based upon work partially supported by the Air Force Office of Scientific Research under award number FA8655-24-1-7009

\section{Appendix: Comparison of $\hat p_c$ and $p_{\rm obs}$} \label{appendix:pc_vs_pvalues}
The MLE of $\xi$ for fixed miss distance $\psi$ is defined as
$$
\hat{\xi}(\psi)= \arg\min_{\|\xi\| = \psi} 
\left\{
\frac{(\xi_1 - x_1)^2}{d_1^2} + \frac{(\xi_2 - x_2)^2}{d_2^2}
\right\},
$$  
so the likelihood root statistic can be expressed as
$$
r(\psi) = \sign(\|x\| - \psi) \left[{\left\{x_1 - \hat{\xi}_1(\psi) \right\}^2}/{d_1^2} + {\left\{x_2 - \hat{\xi}_2(\psi) \right\}^2}/{d_2^2}\right]^{1/2},
$$  
with corresponding \pvalue 
$$
p_{\rm obs}= \Phi\left\{- r(\HBR)\right\} = \dfrac{1}{\sqrt{2\pi}} \int_{r(\HBR)}^{\infty} e^{-t^2/2} dt.
$$  
If we define the function  
$$
\Delta(t) = \frac{(t_1 - x_1)^2}{d_1^2} + \frac{(t_2 - x_2)^2}{d_2^2},
$$  
then we can write  
$$
r(\HBR) = \sign(\|x\| - \HBR)\Delta(\hat{\xi})^{1/2}
$$  
where  $\hat{\xi}= \hat{\xi}(\HBR)$.  To proceed, we define the sets
\begin{flalign*}
A &= \{ t \in \mathbb{R}^2 : \|t\| \leq \HBR \}, \\
C &= \{ t \in \mathbb{R}^2 : \Delta(t) \leq \Delta(\hat{\xi}) \},\\
B &= \{ t\in \mathbb{R}^2 : t_1 \hat{\xi}_1 + t_2 \hat{\xi}_2 \leq \|\hat{\xi}\|^2 \},\\
D &= \left\{ t\in \mathbb{R}^2 : {(t_1 - x_1)( \hat{\xi}_1 - x_1)}/{d_1^2} + {(t_2 - x_2)( \hat{\xi}_2 - x_2)}/{d_2^2} \leq \Delta(\hat{\xi}) \right\}.
\end{flalign*}
    
\noindent\textbf{Lemma 1.}  If $\|x\| > \HBR$, then $B = \overline{D^c}$, whereas if $\|x\| < \HBR$, then $B = D$.
\medskip 

\noindent \textit{Proof.} The disk $A = \{ t : \|t\| \leq \HBR \}$ has a differentiable boundary, on which every supporting half-space is unique. The Cauchy--Schwarz inequality implies that the function $u \mapsto u \cdot \hat{\xi}$ is a supporting linear functional, so $B$ is the unique supporting half-space of $A$ at $\hat\xi$. The elliptical set $C = \{ t : \Delta(t) \leq \Delta(\hat{\xi}) \}$ also has a smooth boundary, and its unique supporting half-space at $\hat\xi$ is $D$.  


When $\|x\| > \HBR$, suppose that $\text{int}(A) \cap \text{int}(C) \neq \emptyset$. Then there exists $t$ such that $\|t\| < \HBR$ and $\Delta(t) < \Delta(\hat{\xi})$. By the intermediate value theorem, there exists a convex combination $s = \alpha t + (1-\alpha)x$ such that $\|s\| = \HBR$. Since $\text{int}(C)$ is convex, we obtain $\Delta(s) < \Delta(\hat{\xi})$, contradicting the definition of $\hat{\xi}$. Hence, $\text{int}(A) \cap \text{int}(C) = \emptyset$. For $\|x\| < \HBR$, a similar argument shows that $C \subseteq A$. In both cases, $A$ and $C$ share the boundary point $\hat{\xi}$.

If $\|x\| > \HBR$, the separating hyperplane theorem implies that $B = \overline{D^c}$. However, if $\|x\| < \HBR$, then $C \subseteq A$ implies that $B = D$.\endex

\noindent\textbf{Theorem 1.} In the setting above,     
$$
\hat p_c = \dfrac{1}{2\pi d_1 d_2} \iint_{t \in A} e^{- \Delta(t)/2}\,\D{t_1}\D{t_2}\leq p_{\rm obs}.
$$  
\medskip
\noindent \textit{Proof.} There are three possibilities: that $\|x\| = \HBR$, $\|x\| > \HBR$ or $\|x\| < \HBR$.

If $\|x\| = \HBR$ then $r(\HBR) = 0$ and therefore $p_{\rm obs} =\Phi(0)= 1/2$. The integral defining $\hat p_c$ is over less than one-half of a bivariate normal density and thus is strictly less than $1/2$. Hence, $\hat p_c < p_{\rm obs}$.

The second possibility is that $\|x\|>\HBR$, in which case $B = \overline{D^c}$ by Lemma~1. Hence 
\begin{eqnarray*}
\hat p_c=\dfrac{1}{2\pi d_1 d_2} \iint_{t \in A} e^{- \Delta(t)/2} \,\D{t_1}\D{t_2} &\leq& \dfrac{1}{2\pi d_1 d_2} \iint_{t \in B} e^{-\Delta(t)/2} \, \D{t_1}\D{t_2},\\ &=& \dfrac{1}{2\pi d_1 d_2} \iint_{t \in \overline{D^c}} e^{-\Delta(t)/2} \,\D{t_1}\D{t_2}.
\end{eqnarray*}

To evaluate this integral, we define new coordinates $s=(s_1, s_2)$ by 
$$\begin{pmatrix}\dfrac{t_1-x_1}{d_1} \\ \dfrac{t_2-x_2}{d_2}\end{pmatrix}=  \Delta(\hat\xi)^{-1/2}\left( \begin{array}{cc}
\dfrac{\widehat{\xi}_1-x_1}{d_1} & -\dfrac{\widehat{\xi}_2-x_2}{d_2} \\
\dfrac{\widehat{\xi}_2-x_2}{d_2} & \dfrac{\widehat{\xi}_1-x_1}{d_1}
\end{array}\right)\begin{pmatrix}s_1\\s_2\end{pmatrix}.
$$
This composite transformation from $(t_1,t_2)$ to $(s_1,s_2)$ both standardizes and rotates the space; the determinant of its Jacobian is $d_1 d_2$. The integration region \[\{t: (t_1 - x_1)( \hat{\xi}_1 - x_1)/{d_1^2} + {(t_2 - x_2)( \hat{\xi}_2 - x_2)}/{d_2^2}\geq\Delta(\hat{\xi})\}\] becomes $\{s: s_1 \geq  \Delta(\hat{\xi})^{1/2}, s_2 \in \mathbb{R} \}$. Hence
\begin{eqnarray*}
\dfrac{1}{2\pi d_1 d_2} \iint_{t \in \overline{D^c}} e^{-\Delta(t)/2} \,\D{t_1}\D{t_2}&=&\dfrac{1}{2\pi}\iint_{s_1 \geq \Delta(\widehat{\xi})^{1/2}} e^{-(s_1^2+s_2^2)/2}\,\D{s_1}\D{s_2}\\
&=&\dfrac{1}{\sqrt{2\pi}}\int_{s_1\geq \Delta(\hat{\xi})^{1/2}} e^{-s_1^2/2}\,\D{s_1},
\end{eqnarray*}
and, using the fact that $r(\mathrm{\HBR})=\Delta(\hat{\xi})^{1/2}$, we therefore have 
$$
\hat p_c \leq \dfrac{1}{\sqrt{2 \pi}} \int_{r(\mathrm{\HBR})}^{\infty} e^{-s^2/2} d s=\Phi\left\{-r(\mathrm{\HBR})\right\}=p_{\mathrm{obs}}.
$$

The third possibility is that $\|x\|<$ \HBR. In this scenario $B=D$ so
\begin{eqnarray*}
\hat p_c=\dfrac{1}{2 \pi d_1 d_2} \iint_{t \in A} e^{- \Delta(t)/2} \,\D{t_1}\D{t_2} \leq \dfrac{1}{2 \pi d_1 d_2} \iint_{t \in B} e^{-\Delta(t)/2} \,\D{t_1}\D{t_2},\\
=\dfrac{1}{2 \pi d_1 d_2} \iint_{t \in D} e^{-\Delta(t)/2} \,\D{t_1}\D{t_2}.
\end{eqnarray*}
In this case, using the same transformation as above, the integration region \[\{t: (t_1 - x_1)( \hat{\xi}_1 - x_1)/{d_1^2} + {(t_2 - x_2)( \hat{\xi}_2 - x_2)}/{d_2^2} \leq \Delta(\hat{\xi})\}\] becomes $\{s: s_1 \leq  \Delta(\hat{\xi})^{1/2}, s_2 \in \mathbb{R} \}$, giving
$$
\dfrac{1}{2\pi d_1 d_2} \iint_{t \in D} e^{-\Delta(t)/2} \,\D{t_1}\D{t_2}=\dfrac{1}{2\pi}\iint_{s_1 \leq \Delta(\widehat{\xi})^{1/2}} e^{-(s_1^2+s_2^2)/2}\,\D{s_1}\D{s_2}=\dfrac{1}{\sqrt{2\pi}}\int_{s_1\leq \Delta(\hat{\xi})^{1/2}} e^{-s_1^2/2}\,\D{s_1}.
$$
For $\|x\|<$ \HBR, we have $r(\mathrm{\HBR})=-\Delta(\hat{\xi})^{1/2}$, so 
$$
\hat p_c \leq \dfrac{1}{\sqrt{2 \pi}} \int_{-\infty}^{-r(\mathrm{\HBR})} e^{-s_1^2/2} \,\D{s_1}=\Phi\left\{-r(\mathrm{\HBR})\right\}=p_{\mathrm{obs}}.
$$
\endex

\clearpage
\newpage
\bibliographystyle{plain}
\bibliography{biblio}

\end{document}